\documentclass[conference]{IEEEtran}
\usepackage{graphicx,subcaption}
\usepackage{amssymb,amsmath,epsfig,citesort,url,fixltx2e}
\usepackage{algorithm,algorithmic}
\usepackage[font=footnotesize]{caption}
\psfull

\IEEEoverridecommandlockouts    % To enable thanks in author tag
\title{\LARGE \bf Viewport-Driven Rate-Distortion Optimized $\mathbf{360^\circ}$ Video Streaming\vspace{-0.5cm}}
\author{Jacob Chakareski\thanks{The work of J. Chakareski and R. Aksu has been supported in part by NSF award CCF-1528030 and a research gift from Adobe Systems.}, Ridvan Aksu, Xavier Corbillon, Gwendal Simon, and Viswanathan Swaminathan}

\begin{document}

\maketitle

%%%%%%%%%%%%%%%%%%%%%%%%%%%%%%%%%%%%%%%%%%%%%%%%%%%%%%%%%%%%%%%%%%%%%%%%%%%%%%%%
\begin{abstract}
The growing popularity of virtual and augmented reality communications and $\mathbf{360^\circ}$ video streaming is moving video communication systems into much more dynamic and resource-limited operating settings. The enormous data volume of $\mathbf{360^\circ}$ videos requires an efficient use of network bandwidth to maintain the desired quality of experience for the end user. To this end, we propose a framework for viewport-driven rate-distortion optimized $\mathbf{360^\circ}$ video streaming that integrates the user view navigation pattern and the spatiotemporal rate-distortion characteristics of the $\mathbf{360^\circ}$ video content to maximize the delivered user quality of experience for the given network/system resources. The framework comprises a methodology for constructing dynamic heat maps that capture the likelihood of navigating different spatial segments of a $\mathbf{360^\circ}$ video over time by the user, an analysis and characterization of its spatiotemporal rate-distortion characteristics that leverage preprocessed spatial tilling of the $\mathbf{360^\circ}$ view sphere, and an optimization problem formulation that characterizes the delivered user quality of experience given the user navigation patterns, $\mathbf{360^\circ}$ video encoding decisions, and the available system/network resources. Our experimental results demonstrate the advantages of our framework over the conventional approach of streaming a monolithic uniformly-encoded $\mathbf{360^\circ}$ video and a state-of-the-art reference method. Considerable video quality gains of 4 - 5 dB are demonstrated in the case of two popular 4K $\mathbf{360^\circ}$ videos.
\end{abstract}

%%%%%%%%%%%%%%%%%%%%%%%%%%%%%%%%%%%%%%%%%%%%%%%%%%%%%%%%%%%%%%%%%%%%%%%%%%%%%%%%
\section{Introduction}
Emerging virtual and augmented reality (VR/AR) technologies are helping introduce novel immersive digital experiences. It is anticipated that VR/AR technologies will represent a \$108 billion market in the near future \cite{digi2017}. Gaming, entertainment, education and training, and $360^\circ$ video are the main application domains of VR/AR technologies at present, with a broader set of societal applications spanning remote sensing, the environmental and weather sciences, disaster relief, and transportation anticipated in the future \cite{ApostolopoulosCCKTW:12}.

\begin{figure}[htb]
\centering
  \includegraphics[width=\columnwidth]{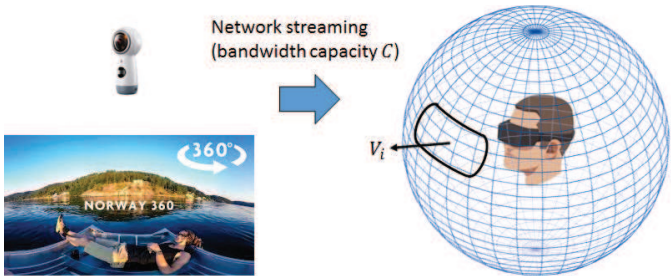}
  \caption{360$^\circ$ streaming: Viewport $V_i$ on the 360$^\circ$ sphere.}
  \label{fig:360ViewSetup}
\end{figure}

360$^\circ$ video is an emerging video format captured by an omnidirectional camera that records incoming light rays from every direction. It enables a 360$^\circ$ look-around of the surrounding scene for a remote user, virtually placed at the camera location, on his VR device, as illustrated in Figure~\ref{fig:360ViewSetup}. Presently, the entire monolithic 360$^\circ$ view panorama is streamed to the user, who, however, can only experience a small portion of it denoted as viewport $V_i$, at any time, as also illustrated in Figure~\ref{fig:360ViewSetup}. However, this results in a huge network overhead/bottleneck and unnecessary computational/bandwidth loading of the device, which, in turn, considerably penalize the user quality of experience. Moreover, to apply conventional video coding, the 360$^\circ$ view sphere is first mapped to a planar shape: equirectangle, pyramid, cube, or dodecahedron. The latter three have been considered since around 30\% pixel replication is introduced when the sphere is mapped to an equirectangle \cite{Corbillon2017,Yu2015}. However, they have their own deficiencies, e.g., introduction of projection distortions around the planar shape's edges. Here, we only consider the equirectangular mapping, as the most widely used. % and only one currently considered by the MPEG VR standardization forum.

The growing popularity of VR/AR technologies stimulates an equivalent increasing demand for 360$^\circ$ video content, which today can be accessed through over-the-top online providers such as YouTube/Facebook 360 \cite{Facebook360,YouTube360}. However, present
360$^\circ$ streaming practices necessitate excessive data rates that even anticipated broadband network access technologies would not be able to support \cite{Forbes_VR_2016,EdwardKnightlyKeynoteINFOCOM2017}, due to the heuristic design shortcomings of the former outlined above. Thus, a broader/faster adoption of emerging 360$^\circ$ technologies that can eventually dominate the market is precluded. On the other hand, delivering the entire 360$^\circ$ view sphere is necessary to avoid simulator/motion sickness \cite{MossM:11} that would degrade the quality of experience, as the {\em intuitive approach of sending only $V_i$} using traditional server-client delivery architectures, where the server responds to client updates, would preclude application interactivity, due to the inherent network round-trip latency.

To overcome this apparent impasse between 360$^\circ$ application requirements and technology capabilities/design, which essentially stems from the direct application to the 360$^\circ$ domain of existing video coding/streaming technologies that treat 360$^\circ$ content as conventional videos, recent studies have considered uneven spatial quality encoding of 360$^\circ$ videos, to minimize the data rate assigned to 360$^\circ$ regions not navigated by the user presently, thereby considerably reducing the induced network overhead. This is the strategy we also follow, making the following contributions in this context.

We formulate a framework for viewport-driven rate optimized $360^\circ$ video streaming that integrates the user view navigation pattern and the spatiotemporal rate-distortion characteristics of the $360^\circ$ video content to maximize the delivered user quality of experience for the given network/system resources. It comprises (i) a methodology for constructing dynamic heat maps that capture the user likelihood of navigating different spatial segments of a $360^\circ$ video over time, (ii) an analysis and characterization of its spatiotemporal rate-distortion characteristics that leverage a preprocessed spatial tilling of the $360^\circ$ view sphere, and (iii) an optimization problem formulation that characterizes the delivered user quality of experience given the user navigation patterns, $360^\circ$ video encoding decisions, and the available system/network resources.

The rest of the paper is organized as follows. In Section~\ref{sec:RelatedWork}, we first review related work. Subsequently, we present the building components of our system framework in Section~\ref{sec:SystemModels}. The problem formulation that aims to maximize the delivered $360^\circ$ user quality of experience given the user navigation patterns, $360^\circ$ video encoding decisions, and the available system/network resources, is presented in Section~\ref{sec:ProblemFormulationOptimization}. Experimental analysis of the performance of our framework and validation of our system models is carried out in Section~\ref{Experiments}. Finally, concluding remarks and a summary of envisioned future work are provided in Section~\ref{Conclusion}.
%\vspace{-0.1cm}

\section{Related Work}
\label{sec:RelatedWork}
%\vspace{-0.1cm}
Due to the emerging nature of $360^\circ$ technologies, only a small body of related work has appeared to date. The study in \cite{Afzal2017} carried out an empirical characterization of $360^\circ$ videos highlighting their main features, e.g., their lower temporal rate variability compared to conventional videos. A small number of studies have considered splitting the $360^\circ$ video into spatial tiles as part of the encoding process, leveraging the tilling feature of the latest High Efficiency Video Coding (HEVC) standard \cite{SullivanOHW:12}. The encoding data rate of each tile can then be controlled independently to reduce the overall bandwidth usage \cite{Ozcinar2017,Petrangeli2017,Qian2016,Hosseini2016}. Transmission aspects of $360^\circ$ HTTP streaming have been explored in \cite{Graf2017,Yahia2017}. Applications of scalable video coding to $360^\circ$ streaming have been studied in \cite{Nasrabadi2017,He2018}. Head movement prediction and the impact of navigation uncertainty have been investigated in \cite{Qian2016,Corbillon2017}. The former study also carries out an empirical analysis of the performance efficiency of the four sphere-to-planar shape projections and investigates the benefits of streaming adaptively multiple $360^\circ$ representations featuring different quality-emphasized spatial regions \cite{Corbillon2017}.

In contrast to the few studies cited above that consider HEVC $360^\circ$ tiling, we employ preprocessed spatial tiles of the $360^\circ$ view panorama, which has several advantages in the form of lower complexity at multiple critical aspects of a server-client $360^\circ$ streaming architecture \cite{Corbillon2017}. Moreover, a formal analysis of the spatiotemporal rate-distortion characteristics of $360^\circ$ tiling that integrates the user navigation patterns and the available network/system resources has not been carried out towards optimal selection of $360^\circ$ encoding and streaming decisions. The framework of our paper aims to fill this gap.

\section{System models}
\label{sec:SystemModels}
\vspace{-0.2cm}

\subsection{Overview}
\begin{figure}[htb]
\centering
\includegraphics[width=0.9\columnwidth]{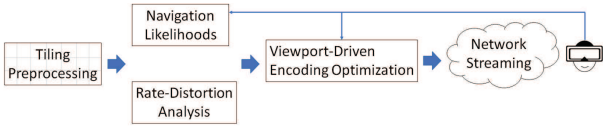}
\caption{$360^\circ$ networked system architecture.}
\vspace{-0.6cm}
\label{sysmodel}
\end{figure}

Our $360^\circ$ networked system architecture comprises several major component blocs that were introduced earlier and is illustrated in Figure~\ref{sysmodel}. We describe each one in detail here.

\subsection{Tiling preprocessing}
\label{sec:TilingPreprocessing}
\begin{figure}[htb]
\centering
\vspace{-0.2cm}
\includegraphics[width=0.9\columnwidth]{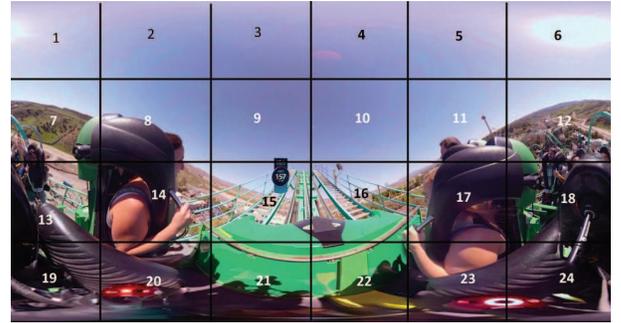}
\caption{$360^\circ$ video panorama $6 \times 4$ spatial tiling.}
\vspace{-0.3cm}
\label{tiles}
\end{figure}

We partition a $360^\circ$ video into a set of $N \times M$ spatial tiles. In particular, we partition the raw $360^\circ$ video frames into spatial tiles and consider the collection of thereby constructed (smaller) video frames for each tile as separate videos. The tiles are then separately encoded and streamed to the user, according to our analysis and optimization. As explained earlier, carrying out tiling as a preprocessing step has several advantages over tiling the video as part of the encoding process, as enabled by the tiling feature of the latest video coding standard HEVC. In our experiments, we used two popular 4K $360^\circ$ videos \cite{Coaster,Dubai} that we preprocessed into $6 \times 4$ spatial times, as illustrated in Figure~\ref{tiles}, where the first and second dimension refer to the horizontal and vertical pixel resolution of the video. Each tile is indexed in a raster fashion, top-to-bottom and left-to-right.

We selected this specific tiling based on empirical analysis, as a reasonable choice between the complexity and compression efficiency introduced by a given tiling. In a follow-up study, we plan to integrate the selection of tiling as part of the $360^\circ$ end-to-end analysis and optimization.

\subsection{$360^\circ$ VR head movement data}
\begin{figure}[htb]
\centering
\vspace{-0.2cm}
\includegraphics[width=0.9\columnwidth]{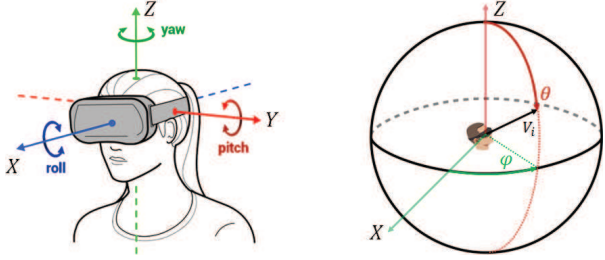}
\caption{$360^\circ$ head movement navigation data of current viewport $V_i$. Left: Rotation angles yaw, pitch, and roll around the three coordinate axis. Right: Azimuthal and polar angles $(\varphi,\theta)$ in spherical coordinates.}
\vspace{-0.5cm}
\label{NavigationData}
\end{figure}

We collected head-movement data that describes how a user navigates a $360^\circ$ video over time. In particular, a VR device outputs the direction of the current viewpoint of the user $V_i$ on the $360^\circ$ view sphere up to 250 times per second, with the user considered to be placed at the sphere center, as described earlier. Precisely, this is the surface normal of $V_i$ on the $360^\circ$ sphere that is uniquely described by the spherical coordinates azimuth and polar angles $\varphi \in [0^\circ,360^\circ]$ and $\theta \in [0^\circ,180^\circ]$ it spans on the sphere, in a spherical coordinate system with the $360^\circ$ sphere center as its origin, as illustrated in Figure~\ref{NavigationData} (right). These two angles are equivalently denoted as yaw and pitch in the VR community, captured as rotation angles around the $Z$ and $Y$ axes, as denoted in Figure~\ref{NavigationData} (left). We collected the pairs $(\varphi_j,\theta_j)$ that coincided with the discrete temporal instances $t_j$ of subsequent $360^\circ$ video frames $j$ displayed to the user as he navigates the content. They are the navigation data points relevant for our analysis.

\subsection{Navigation likelihoods}
\label{sec:NavigationLikelihoods}
\vspace{-0.05cm}

For various head-mounted displays (HMD) used in VR applications, the viewport size experienced by the user varies. In this paper, we assume a viewport of 110$^{\circ}$ horizontal and 90$^{\circ}$ vertical fields of view. For every navigation trace for a given $360^\circ$ video, we compute the fraction of the surface area of tile $k$ occupied by the user viewport $V_i$ at time instance $j$, denoted as $w_{k,j}$. To account for the unequal surface area occupied by different viewports, when mapped to a 2D rectangle used to encode the data, depending on their latitude (polar angle $\theta$) on the $360^\circ$ view sphere, each tile $k$ is assigned a normalized weight $\bar{w}_{k,j}$, computed as $\bar{w}_{k,j} = w_{k,j}/\sum_k w_{k,j}$. We can then aggregate these weights over different time durations, to compute the likelihoods of navigating different tiles of the respective $360^\circ$ video during those time periods. In our analysis, we are interested in exploiting these navigation likelihoods over the duration of individual Groups Of Pictures (GOPs) comprising the encoded $360^\circ$ content.

\begin{figure}[htb]
\centering
\vspace{-0.2cm}
\includegraphics[width=0.85\columnwidth]{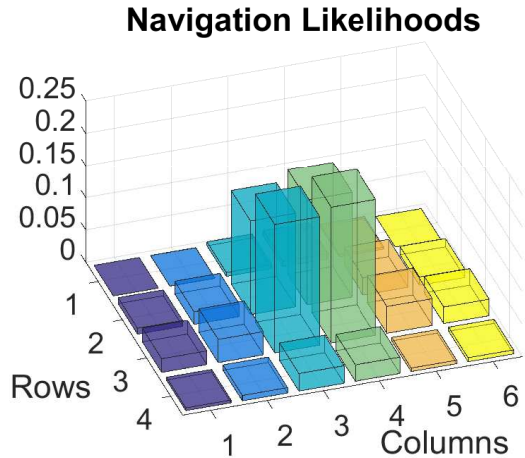}
\caption{Navigation likelihoods of tiles for Roller Coaster.}
\vspace{-0.2cm}
\label{weight}
\end{figure}

For illustration, Figure~\ref{weight} shows the average (over the entire video) navigation likelihoods of different tiles comprising the selected $6 \times 4$ tiling applied to the $360^\circ$ video Roller Coaster used in our experiments. We can see that corner tiles appear rarely in a viewport navigated by the user, as their navigation likelihoods are close to zero. Conversely, it appears that the user often navigated through tiles 9, 10, 15, and 16, for instance, as they have much higher navigation likelihoods.

%
%that would mostly appeared within the viewport usually would have a very high weight eg. tiles 9, 10, 15, and 16.
%
%
%Figure \ref{weight2}
%
%and Wingsuit.
%
%
%in the video content over that time period
%
%
%
%
%
%HMD devices that are on the market, the viewport size varies. In this paper a viewport of 110$^{\circ}$ horizontal and 90$^{\circ}$ vertical is considered as done by Qian et al.~\cite{Qian2016}. Depending on the area of the tile within the viewport for each frame, each tile gets a weight within interval (0,1). Extending this for all frames within the chunk, we obtain the weight of each tile. And for the tiles that have never appeared within the viewport throughout the history of that chunk, that GOP is going to be skipped for the tile.
%
%Figure \ref{weight} shows the normalized weight of each tile throughout the Roller Coaster video~\cite{Coaster}. In this video, tiles that would appear rarely within the viewport would have weight close to zero eg. tiles in the corners. Conversely, tiles that would mostly appeared within the viewport usually would have a very high weight eg. tiles 9, 10, 15, and 16.

%In the second video that we have used, viewport is mostly on lower tiles due to video content ~\cite{Dubai}. This video has smoother distribution of weights and focus is mostly on tiles 13-24.

\begin{figure}[htb]
\centering
\vspace{-0.3cm}
\includegraphics[width=0.85\columnwidth]{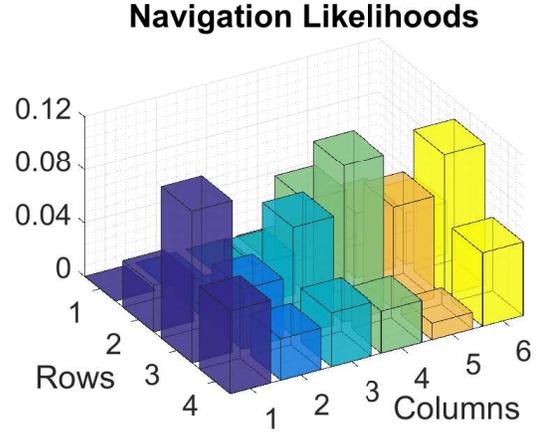}
\vspace{-0.2cm}
\caption{Navigation likelihoods of tiles for Wingsuit.}
\label{weight2}
\vspace{-0.2cm}
\end{figure}

Figure~\ref{weight2} shows the corresponding tile navigation likelihoods for the second $360^\circ$ video, Wingsuit, used in our experiments. It appears that in this case the viewport navigated by the user is mostly closer to the south pole, as the corresponding tiles have much higher likelihoods now, due to the specific nature of this video (more interesting content is spatially located there).

%In the second video that we have used, viewport is mostly on the south pole tiles due to video content~\cite{Dubai}. Contrary to the first video, focus is on tiles in lower part of the video, where the content is more interesting (Figure \ref{weight2}).

%For a more realistic case viewing history of the video can be used as a guide. This allows a model for the tiles that are not viewed but has a high probability to appear in the FoV.

%\begin{figure}
%\centering
%\includegraphics[width=0.9\columnwidth]{Figures/viewport.eps}
%\caption{Navigation of viewport.}
%\label{navi}
%\end{figure}

\begin{figure}[htb]
\centering
\vspace{-0.2cm}
\includegraphics[width=0.9\columnwidth]{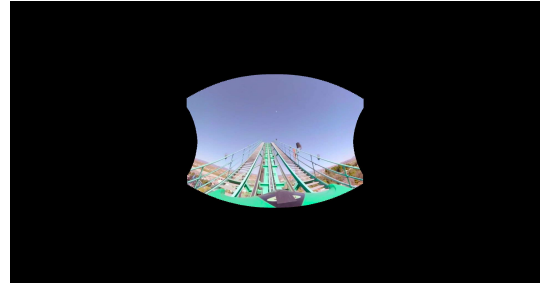}
\vspace{-0.2cm}
\caption{Viewport at $(\varphi,\theta) = (0^{\circ},0^{\circ})$.}
\vspace{-0.6cm}
\label{vp1}
\end{figure}

\begin{figure}[htb]
\centering
\includegraphics[width=0.9\columnwidth]{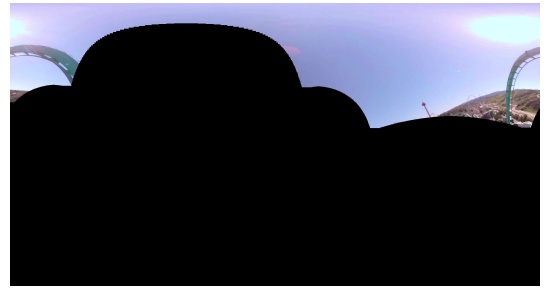}
\vspace{-0.2cm}
\caption{Viewport at $(\varphi,\theta) = (120^{\circ},-60^{\circ})$.}
\vspace{-0.2cm}
\label{vp2}
\end{figure}

A visualization of two representative viewports is shown in Figures \ref{vp1} and \ref{vp2}. Since mapping a 3D shape (sphere) to 2D causes distortion, the shape of a viewport also changes. In equatorial regions, a viewport is smaller and more compact (Figure \ref{vp1}) while in polar regions a viewport is spread over all polar tiles (Figure \ref{vp2}). Figure~\ref{tiles} can be referenced to understand the spatial locations of these two viewports relative to the underlying tiling of the respective $360^\circ$ video.

\subsection{Rate-Distortion models}
\label{sec:r-dModels}
Changing the quality of tiles is a useful method to control the bitrate of a $360^\circ$ video. The quantization parameter QP employed by the HEVC (H.265) codec is a convenient tool for tile quality adaptation. We explore two prospective characterizations of the dependency between the parameter QP and the resulting bitrate $R$ of the encoded tile. That is, we investigate modeling $R = f(\text{QP})$ via an exponential or power law function for $f$ as follows

\begin{equation}
R = a_1 e^{-b_1 \, \text{QP}} \quad \text{or} \quad R = a_2 \text{QP}^{b_2}.
\end{equation}

We will validate these relationships by comparing the bitrate and QP for en encoded $360^\circ$ tile in Section~\ref{sec:ModelValidation}. Since we have a function between the bitrate and QP, we can define bounds for our optimization problem with the highest and lowest QP values that can be selected. And after calculating the optimal bandwidth, going back to QP value and encoding the tiles accordingly can be done easily in the server side.

Similarly, we investigate two prospective characterizations of the dependency between the encoded tile bitrate $R$ and the induced reconstruction error or distortion $D$ for a tile, where the latter can be calculated as the mean-square error (MSE) between the encoded tile video data and the corresponding raw video data for the tile. In essence, the distortion $D$ captures the average deviation of encoded tile pixels from their raw data counterparts. In a raw $360^\circ$ YUV 4:2:0 video, for every pixel sample of the color (chrominance) components U and V there are 4 pixel samples of the (monochromatic) intensity (luminance) component $Y$. Thus, the luminance distortion dominates the encoding distortion for the two color components. Therefore, we used the luminance component distortion as the representative of the encoding distortion $D$ for a tile, measured for every $360^\circ$ tile luminance video frame. We investigate modeling the dependency $D = f(R)$ via an exponential or power law function for $f$ as follows

\begin{equation}
D = c_1 e^{-d_1 R} \quad \text{or} \quad D = c_2 R^{d_2}.
\end{equation}

We validate these relationships by comparing the encoding bitrate and distortion for a encoded $360^\circ$ tile in Section~\ref{sec:ModelValidation}. The characterizations $R = f_1(\text{QP})$ and $D = f_2(R)$ will allow us to formulate the aggregate $360^\circ$ video encoding quality and pursue related optimizations, as explained next.

\section{Optimization Framework}
\label{sec:ProblemFormulationOptimization}
Given the analytical modeling of the relevant problem variables, we now set out to find the optimal bitrate for each tile. There are constraints that we integrate into the problem formulation. These are the aggregate available network bandwidth $C$ and the allowed QP range per tile.

\subsection{Problem Setup}
\label{sec:ProblemFormulation}
Given the limited network bandwidth, tiles should be transmitted at data rates according to their navigation likelihoods and rate-distortion characteristics such that we can maximize the delivered aggregate quality of the respective $360^\circ$ video. Let $R_i(\text{QP}_i)$ denote the bitrate of the i\textsuperscript{th} tile where QP is the encoding quantization parameter, as introduced earlier. This gives us the following inequality to maintain:
\begin{equation}
\sum_i R_i(\text{QP}_i) \leq C, \; i = 1, \ldots, M \times N.
\end{equation}

For practical reasons, for every tile $i$ we set a range of QP values that can be considered, defined by the upper and lower bounds $\text{QP}_{\min}$ and $\text{QP}_{\max}$. This therefore induces constraints on the minimum and maximum data rates that can be assigned to a tile, given the monotonic relationship between QP and $R_i$, as captured by the function $R_i(\text{QP})$. Formally, these two constraints can be written as
\begin{equation}
R_i(\text{QP}_{\max}) \leq R_i(\text{QP}_i) \leq R_i(\text{QP}_{\min}).
\end{equation}

Finally, we formulate the expected 360$^\circ$ quality of experience that a user observes while navigating the scene, as the navigation likelihood weighted sum of video qualities of all tiles comprising the 360$^\circ$ video content streamed to the user. This can be formally written as $\sum_i p(i | v) D_i(R_i)$, where $p(i | v)$ denotes the navigation likelihood of tile $i$ given that viewport $v$ is requested initially. To be precise, note that we formulated our objective as the expected 360$^\circ$ video distortion, due to the one-to-one correspondence between video quality and reconstruction error (distortion). Therefore, we aim to minimize our objective function, as it will lead to the same goal (maximum 360$^\circ$ quality of experience).

\subsection{Optimization Formulation}
Leveraging the problem setup described earlier, we can now formulate the optimization problem of interest as
%\begin{equation}
\begin{align}
& \min_{\{R_i\}}
& & \sum_i p(i|v) D_i(R_i), \label{eqn:Optimization}\\
& \text{subject to:}
& & \sum_i R_i(\text{QP}_i) \leq C, \; i = 1, \ldots, M \times N, \nonumber \\
&
& & R_i(\text{QP}_{\max}) \leq R_i(\text{QP}_i) \leq R_i(\text{QP}_{\min}), \, \forall i. \nonumber
\end{align}

Note that \eqref{eqn:Optimization} represents a convex optimization problem, due to the nature of the constraints involved and the objective function under consideration. Therefore, it can be efficiently solved using fast convex optimization methods \cite{BoydV:03}. In our experiments, we carry out the optimization in \eqref{eqn:Optimization} for every GOP, facilitating the dynamic weight assignment described in Section~\ref{sec:NavigationLikelihoods} to compute the navigation likelihoods $p(i|v)$. In particular, after the optimization completes, the QP vs. $R$ dependency for each tile $i$ in a GOP is used to obtain the explicit optimal $\text{QP}_i$ value that corresponds to the optimal data rate $R_i^*$ produced by \eqref{eqn:Optimization}. Note that for illustration we included the average navigation likelihoods $p(i|v)$ across the applied $360^\circ$ video tilling for the duration of the entire video in Figures \ref{weight} and \ref{weight2}, for the $360^\circ$ content used in our experiments.

We recall that the analytical dependencies between $R$ and $D$, and between $R$ and QP are not explicitly denoted in \eqref{eqn:Optimization}. As explained earlier, we explore two models for each dependency $D=f(R)$ and $R=f(\text{QP})$, an exponential one and a power-law one. And the parameters that comprise each model are extracted uniquely for each tile, before we carry out the optimization in \eqref{eqn:Optimization}. In our experiments, we first validate each of these models, for each dependency, and select the one that is more accurate, to carry out the remaining performance evaluation analysis.

%Bandwidth limitation and the QP bounds are the constraints for the optimization. First one is the external constraint by the system. Second one is the limits for minimum and maximum quality for the tiles. It should be noted that \eqref{eqn:Optimization} represents a convex optimization, thus it can be solved efficiently. After the optimization is completed, the QP vs $R$ dependency for each tile in that segment/GOP is used to obtain the explicit optimal QP values that should be applied to encode every tile and that correspond to the optimal encoding rates $R_i^*$ produced by the optimization.

\section{Experimentation}
\label{Experiments}

\subsection{System Setup}
\label{ssec:experimentsetup}
%In our test system we have used a  4K videos from Youtube~\cite{Coaster,Dubai}. Each video divided into 6 $\times$ 4 tiles and each tile has divided into individual chunks where each chunk is composed of one GoP. Test users have watched those videos with Head Mounted Display devices and their movement has been tracked using OpenTrack software~\cite{OpenTrack}. Based on those movements a heat map is generated for each chunk that shows which tiles have been present within the tile with which percentage (Figures \ref{weight},~\ref{weight2}). Using the heat map, weight of each tile is calculated for the chunk.
%
%For each tile, distortion and bitrate coefficients are extracted for 5 QP values between 22-42. It is stated that distortion vs bitrate is an exponential function and bitrate vs QP is a power law function. Experimental proof of those functions are shown in the following subsection (IV-B). Since all the constraints are provided, optimization problem in (4) can be solved using Matlab. It is assumed that 5 Mbps network speed should be sufficient for a 4K video with a mediocre overall quality.
%
%Results of the optimization function gives the bitrate for each tile. Based on those bitrates, QP for each tile can be defined. To evaluate the quality of the video, QoE throughout the video can be found by frame by frame comparison of FoVs of the encoded video and original video.

We used two popular 4K 360$^\circ$ videos from Youtube, Roller Coaster~\cite{Coaster} and Wingsuit \cite{Dubai}, to evaluate the performance of our framework. VR users watched these videos with HMD devices and their head movements have been tracked using the OpenTrack software~\cite{OpenTrack}. For performance evaluation, we have used one head movement trace per user per video. Based on the collected traces, the navigation likelihoods of each tile in a GOP are calculated, as discussed in Section~\ref{sec:NavigationLikelihoods}.

Each 360$^\circ$ video is preprocessed into 6$\times$4 tiles, as explained in Section~\ref{sec:TilingPreprocessing}. Each tile is encoded into Groups of Pictures (GOPs) of size 32 frames using HEVC. There are 60 GOPs in each tile video, corresponding to 1920 frames and 64 seconds of duration of time, assuming a frame rate of 30 fps. Each GOP is encoded using 5 QP values (22, 27, 32, 37, 42). Using the encoded tiles of those QP values, $R-D$ and QP-$R$ parameters are extracted for all tiles and GOPs, to explore the proposed rate-distortion modeling from Section~\ref{sec:r-dModels}.

Two reference methods are examined to compare against. First, an entire monolithic 360$^\circ$ video is encoded using the following 5 QP values (32, 34, 36, 39, 42). In each case, the induced average data rates for every GOP are used as the network bandwidth constraint $C$ in our own optimization in Section~\ref{sec:ProblemFormulation}. Similarly, we also implemented a state-of-the-art method proposed by Petrangeli et al. \cite{Petrangeli2017}. It predicts future viewports accessed by the user, based on the speed and the position of the HMD. Tiles within the current/future predicted viewports in a GOP are encoded with the highest possible QP value. The remaining tiles are encoded with the lowest possible QP value given the remaining bandwidth budget.

\subsection{Rate-distortion model validation}
\label{sec:ModelValidation}
We formulated two prospective models for the dependencies $D=f(R)$ and $R=f(\text{QP})$, described in Section~\ref{sec:r-dModels}. Here, we explore their accuracy in characterizing the encoded 360$^\circ$ video content we considered in our experiments.

\begin{figure}[htb]
\centering
\begin{subfigure}{.5\columnwidth}
  \centering
  \includegraphics[width=1\linewidth]{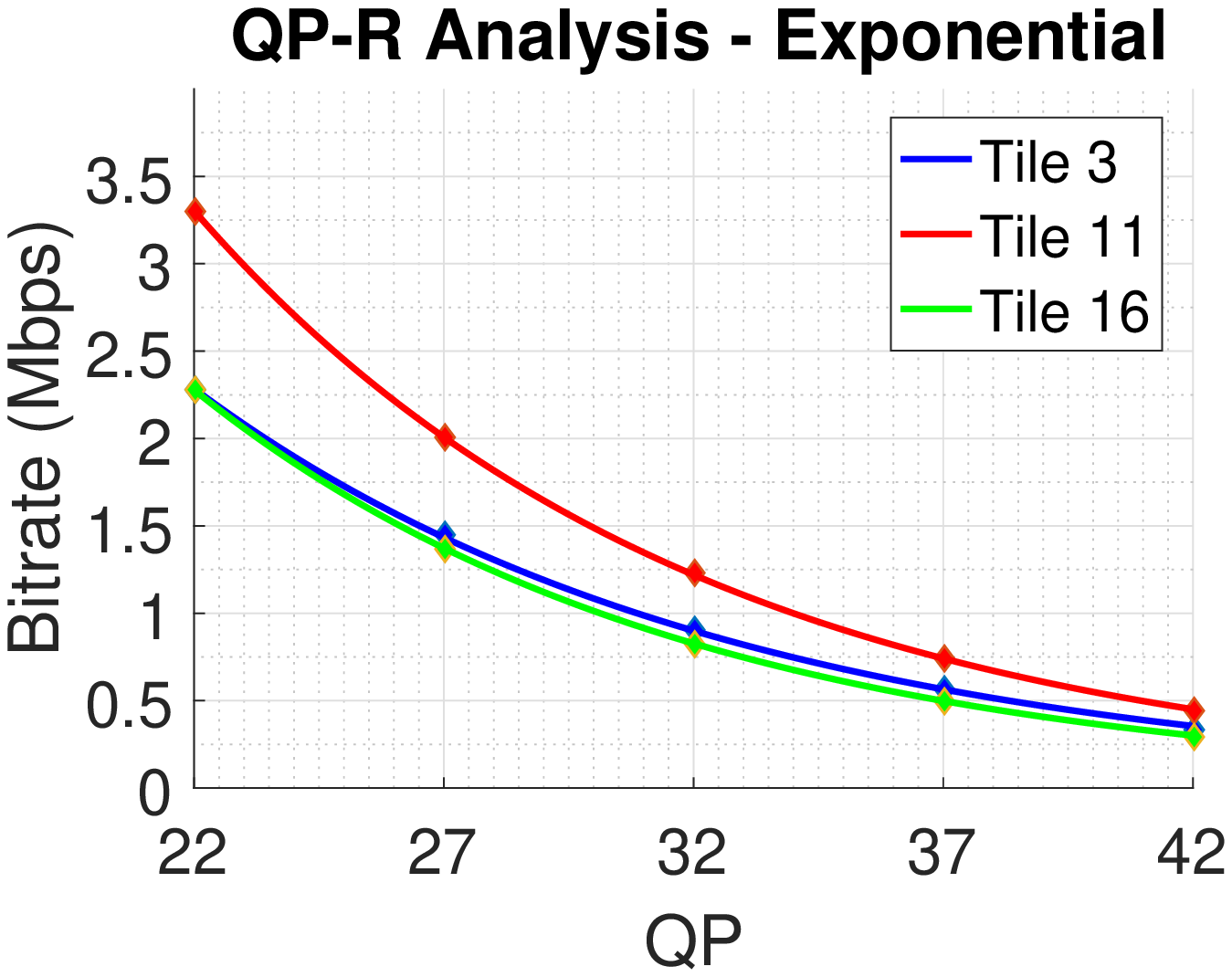}
  \caption{QP vs. bitrate dependency using an exponential model.}
  \label{qp-b-e}
\end{subfigure}%
\begin{subfigure}{.5\columnwidth}
  \centering
  \includegraphics[width=1\linewidth]{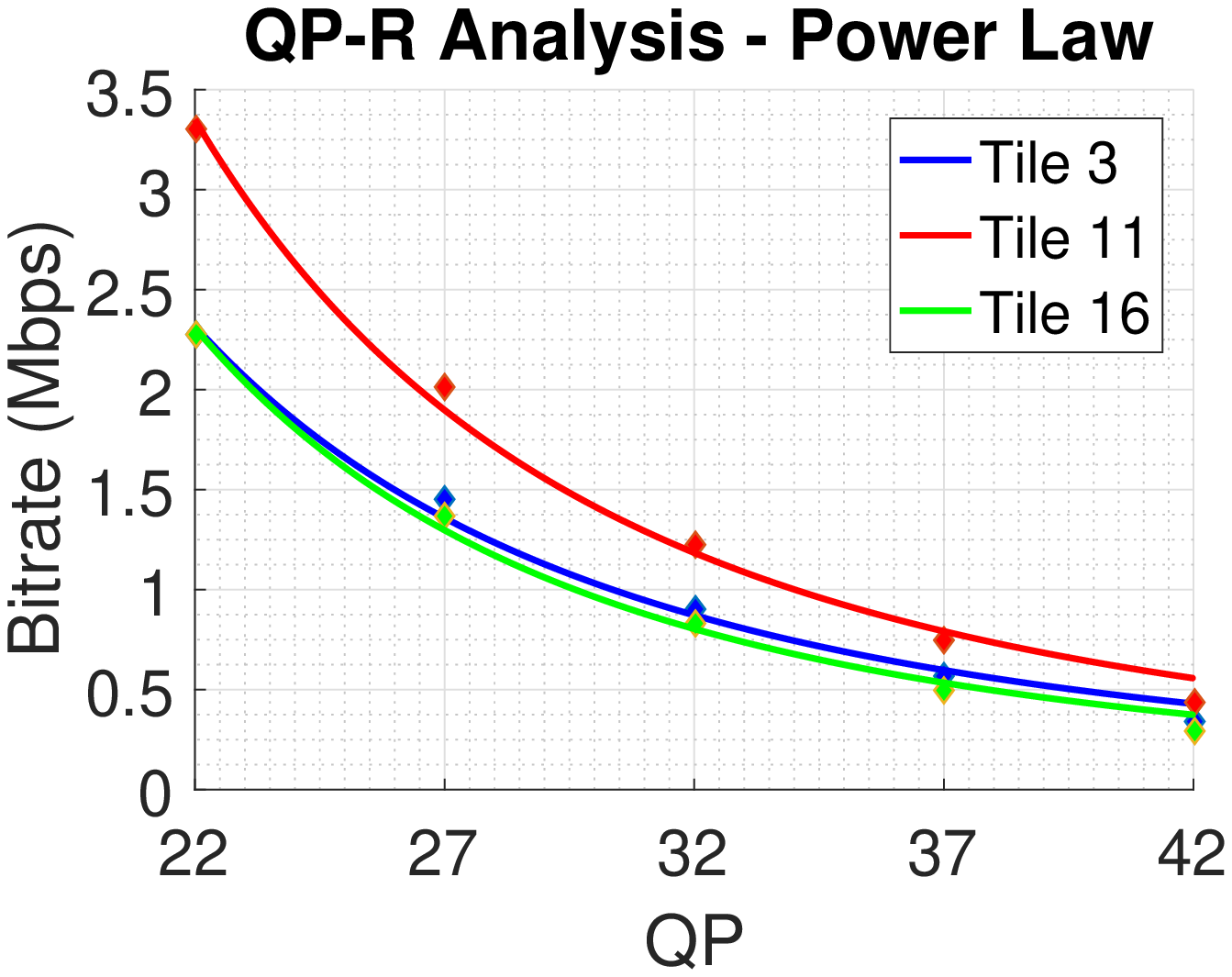}
  \caption{QP vs. bitrate dependency using a power law model.}
  \label{qp-b-p}
\end{subfigure}
\caption{QP vs. bitrate dependency for different tiles. Actual data points shown as markers.}
\label{qp-b}
\vspace{-0.5cm}
\end{figure}

Examination of the employed QP versus induced bitrate relationship for different tiles shows that exponential model fits better the actual data points. In Figure~\ref{qp-b}, we examine these data points, shown as markers, and the fitted analytical dependencies according to the two formulated models, for three representative tiles, with diverse rate-distortion characteristics, from the Roller Coaster video. Referencing the tile indexing from Figure~\ref{tiles}, we can see that while tiles 3 and 16 show lower bitrate requirements due to their relatively static nature, tile 11 requires a higher bitrate as it corresponds to a more dynamic 360$^\circ$ region.

%For bitrate and distortion function, same comparisons have been made. Again red line is the tile 16 from Figure \ref{tiles}, while blue line is tile 11 and green line is tile 3. Middle tile requires high bitrate since it has lots of changes in pixels while others stay similar and needs less bitrate. Also they all fits the power law model well (Figure \ref{b-d}).

\begin{figure}[htb]
\centering
\begin{subfigure}{.5\columnwidth}
  \centering
  \includegraphics[width=1\linewidth]{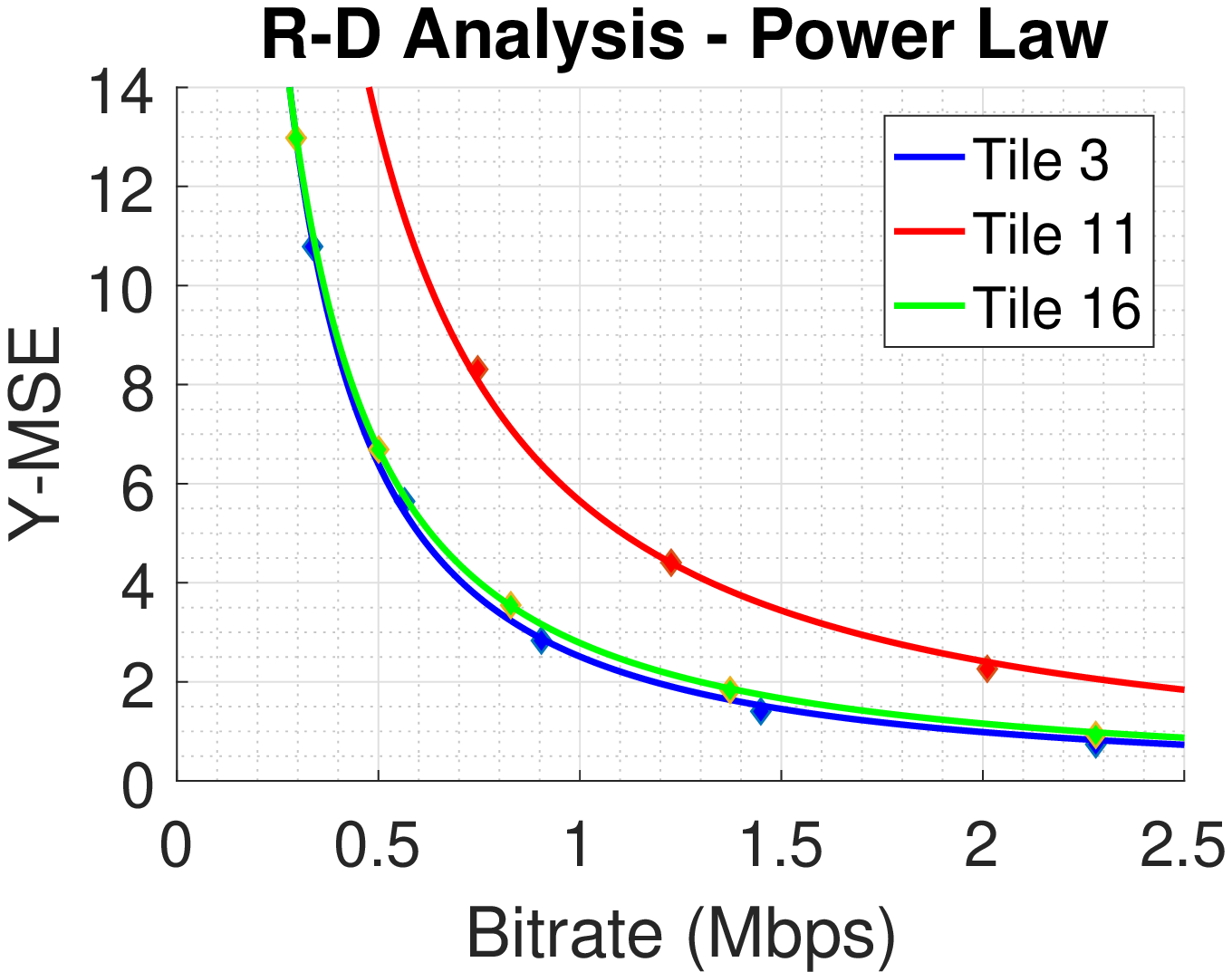}
  \caption{Bitrate vs. distortion dependency using a power law model.}
  \label{b-d-p}
\end{subfigure}%
\begin{subfigure}{.5\columnwidth}
  \centering
  \includegraphics[width=1\linewidth]{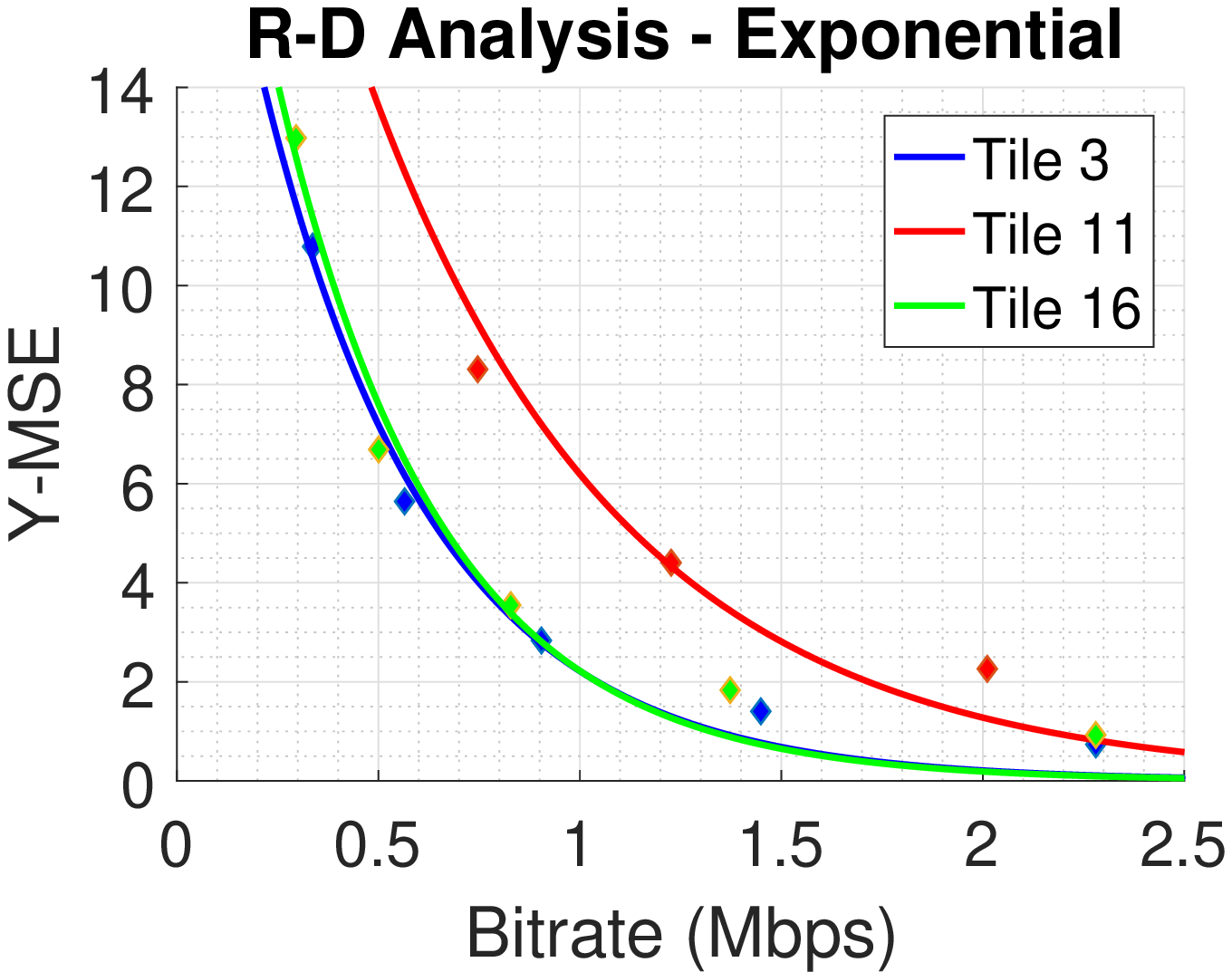}
  \caption{Bitrate vs. distortion dependency using an exponential model.}
  \label{b-d-e}
\end{subfigure}
\caption{Bitrate vs. distortion dependency for different tiles. Actual data points shown as markers.}
\label{b-d}
\vspace{-0.5cm}
\end{figure}

Figure~\ref{b-d} shows the advantage of the power law model in describing the observed $D$ versus $R$ dependency, denoted with markers, across the 360$^\circ$ video tiles. In particular, for lower bitrates, the impact of higher distortion dominates for tiles with more dynamic content (Tile 11), while for higher bitrates the difference across differen tiles in this regard becomes smaller, as seen from Figure~\ref{b-d}.

\subsection{Optimal tile QP and data rates vs. available bandwidth}
%Optimal QP values of the tiles are generated based on bandwidth limitation. In a varying bandwidth scenario those optimal values also vary. We observed the effect of different bandwidth values on a GoP. Figure \ref{br-bw} shows an increasing values on the bitrate of the tiles with increasing bandwidth within a chunk.

\begin{figure}[htb]
\centering
\begin{subfigure}{0.5\columnwidth}
  \centering
  \includegraphics[width=0.95\linewidth]{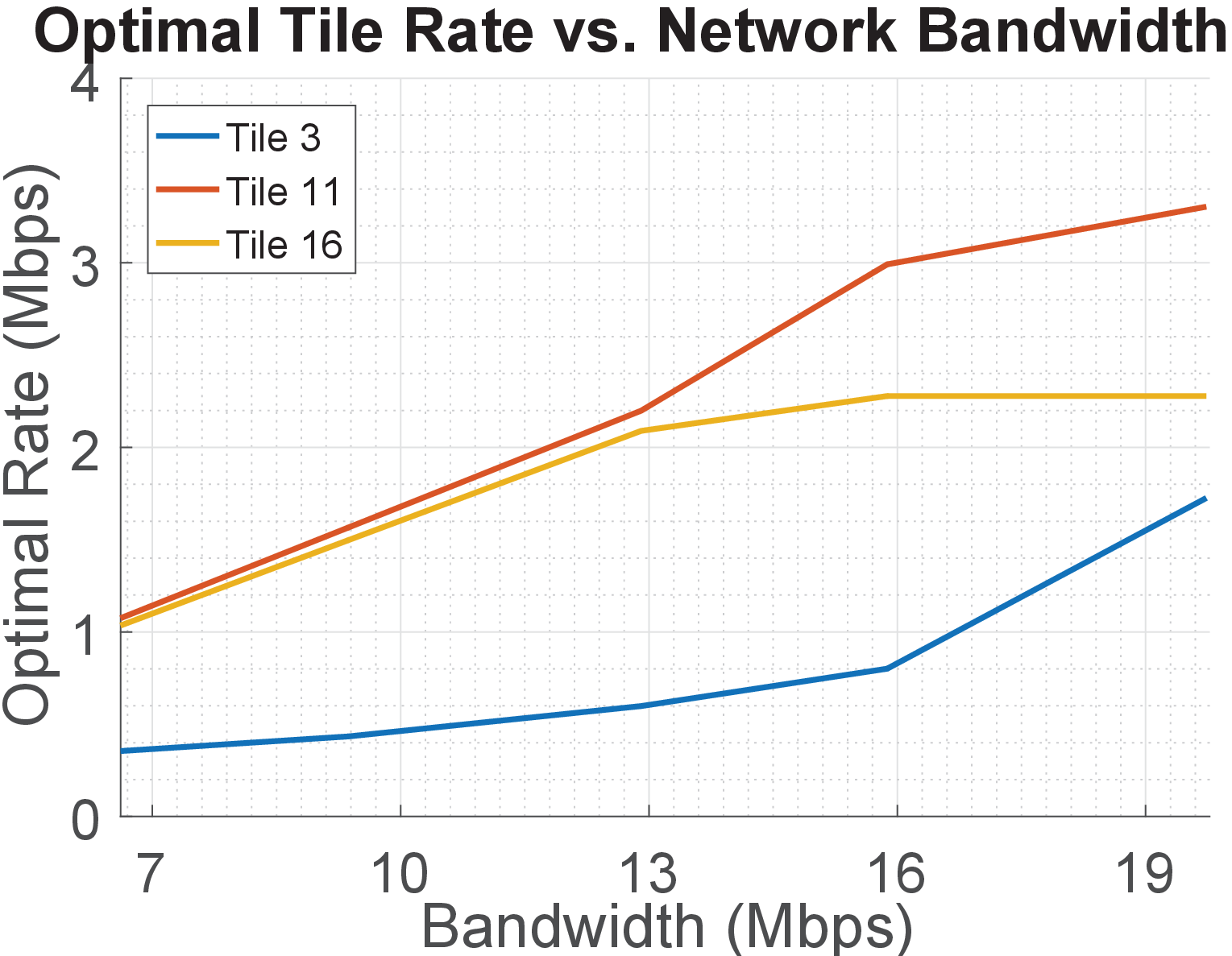}
  \caption{Tile rate vs. bandwidth.}
  \label{br-bw}
\end{subfigure}%
\begin{subfigure}{.5\columnwidth}
  \centering
  \includegraphics[width=1\linewidth]{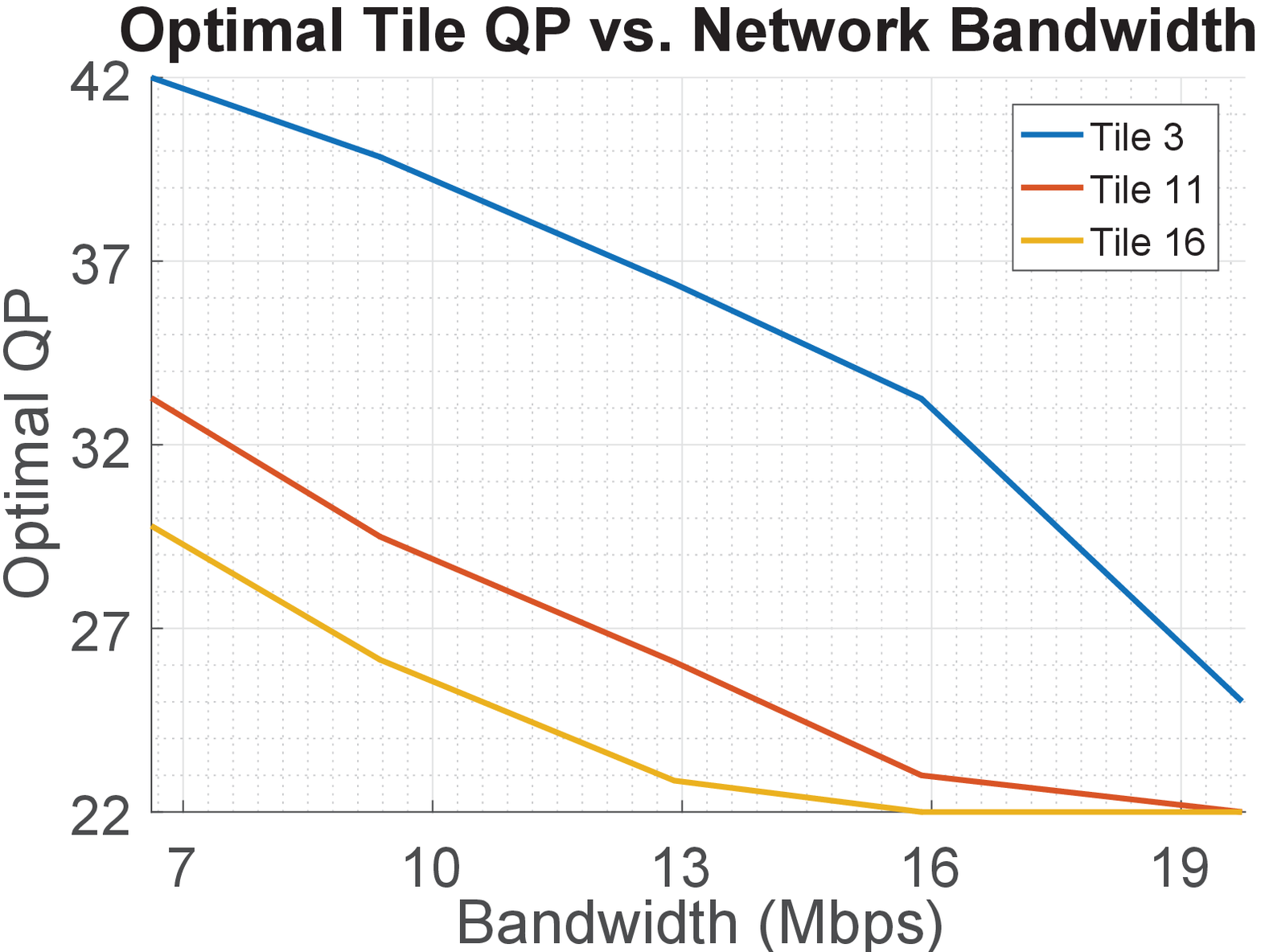}
  \caption{Tile QP vs. bandwidth.}
  \label{qp-bw}
\end{subfigure}
\caption{Optimal tile rate/QP values vs. network bandwidth $C$.}
\label{bw}
\vspace{-0.5cm}
\end{figure}

We examine how the optimal data rates $R_i$ and the corresponding $QP_i$ values, produced by the optimization in \eqref{eqn:Optimization} for every tile $i$, vary, as the available network bandwidth $C$ is varied. Figure~\ref{br-bw} shows the optimal rates produced by \eqref{eqn:Optimization} for three tiles from the Roller Coaster video, for the GOP number 57 in the 360$^\circ$ video, selected as a representative example. For this GOP, tile 3 has a small navigation likelihood, while tile 16 has the highest among the three tiles considered. Still, it is interesting to note that although tile 16 has a higher navigation likelihood relative to tile 11 and is assigned a smaller QP earlier (as seen from Figure~\ref{qp-bw} right), encoding tile 11 leads to a higher data rate in the second half of the graph in Figure~\ref{br-bw}, due to its more dynamic content, which makes encoding it more challenging.

\begin{figure}[htb]
\centering
\vspace{-0.2cm}
\begin{subfigure}{.5\columnwidth}
  \centering
  \includegraphics[width=0.98\linewidth]{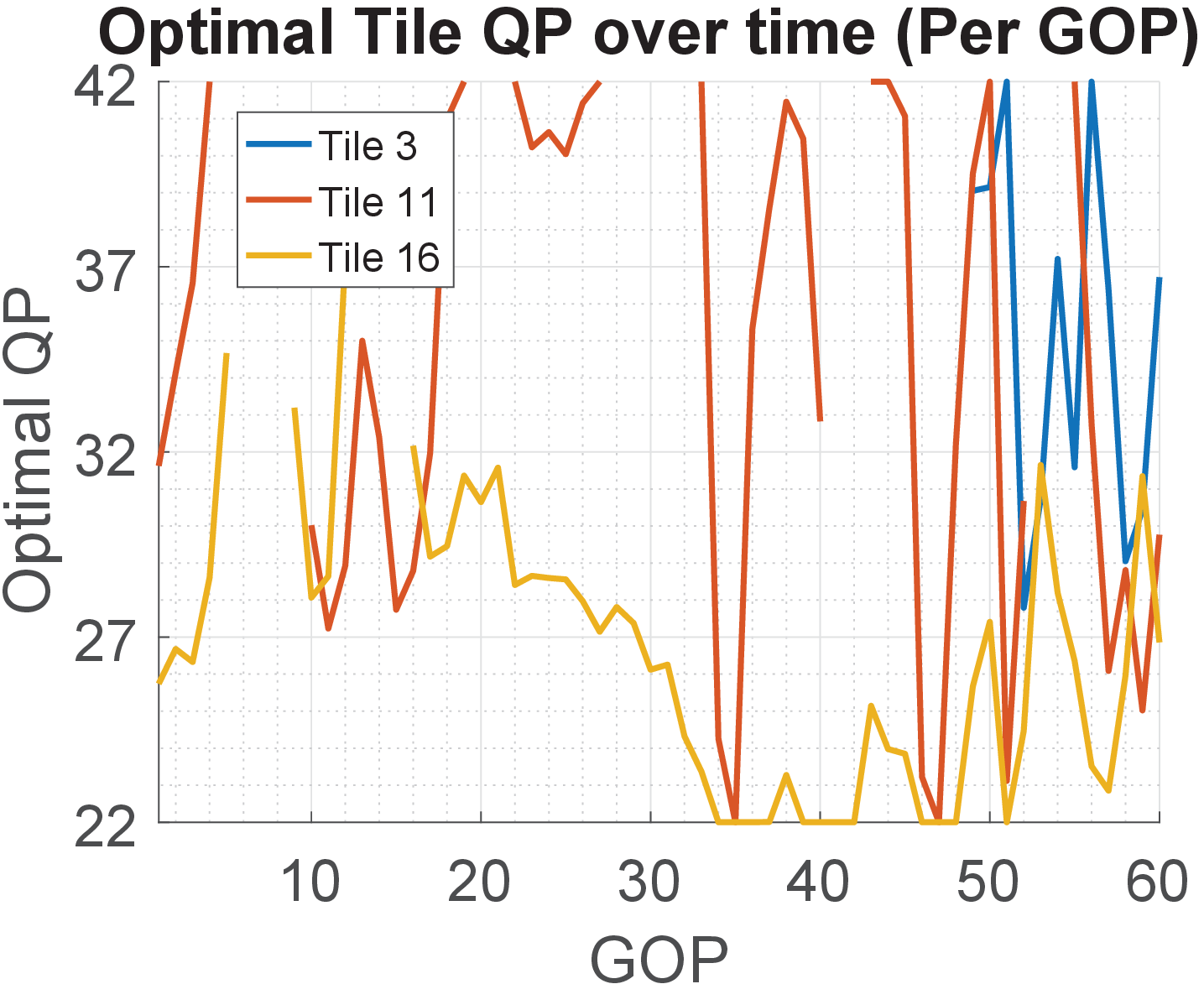}
  \vspace{-0.4cm}
  \caption{Tile QP vs. time.}
  \label{qp-gop}
\end{subfigure}%
\begin{subfigure}{.5\columnwidth}
  \centering
  \includegraphics[width=1\linewidth]{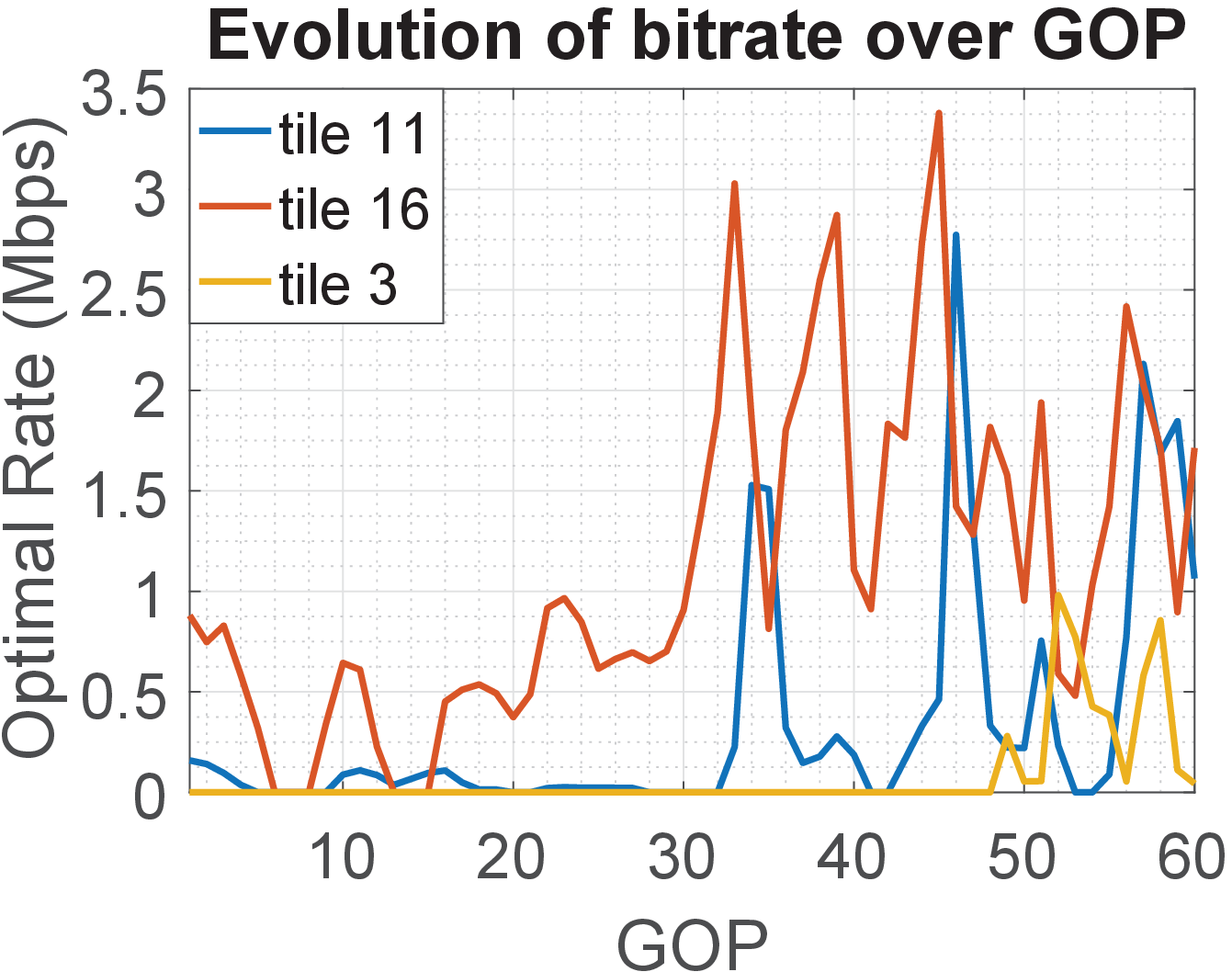}
  \vspace{-0.4cm}
  \caption{Tile rate vs. time.}
  \label{br-gop}
\end{subfigure}
%\vspace{-0.05cm}
\caption{Optimal tile rate/QP values vs. GOP index.}
\label{gop}
\vspace{-0.4cm}
\end{figure}

Figure~\ref{gop} shows the temporal evolution of the optimal QP and bitrate values for these three tiles over the GOPs comprising the 360$^\circ$ content. We can see that tile 16 typically has a lower QP value relative to the other two tiles, due to its frequently accessed spatial location, while tile 3 is often navigated only for a brief period of time towards the end of the video. Discontinuities in Figure~\ref{qp-gop} indicate that a tile has not been assigned any rate (skip encoding mode) by the optimization in \eqref{eqn:Optimization}, as indicated by the corresponding graphs in Figure~\ref{br-gop}.

%In Figure \ref{qp-gop} red line represents  the evolution of QP for tile 16. Due to its central position and high bitrate requirement it usually gets smaller QP values. Discontinuities in \ref{qp-gop} means that tile is not in viewport to be assigned a QP. Tile 16 usually has small QP value except some ripples. Those ripples imply user moved the HMD and weight of that tile is decreased. In Figure \ref{br-gop} optimal bitrate comparison of tiles 3, 11, and 16 can be seen. When the tile is not in viewport, it is not transmitted and has no bitrate.

\subsection{Expected 360$^\circ$ video quality}
%Main point of this work is to provide a fine QoE of 360 video with a sustainable bandwidth. In order to make a realistic example, we have chosen 4K 360 videos and an average bandwidth of 6 Mbps. Crucial question in this sense is to provide a realistic metric in terms of QoE. Although QoE can be thought as a subjective matter which is unique for experience of each user, here we considered the distortion of the viewport of both tiled video and reference video where the constant-QP is used for reference video. In this case, based on the bitrate of the reference video, a bandwidth is calculated and that same bandwidth is used for tiled video.

%The reason we only consider the distortion within the viewport is that the only quality that matters to user is the quality of the tiles within the viewport. In Figure \ref{QoE} green line is showing PSNR of reference system for video 1 where there is no tiling is applied. That leads to a uniform QP for all tiles. The red line shows the PSNR of proposed system. Even though it demonstrate some ripples, most of those ripples stay in a safe boundary. And with an average bandwidth of 6 Mbps, it is possible to watch a 4K 360 video with high quality.

In the analysis here, {\em Proposed} denotes our optimization framework, while {\em Speed-based} and {\em Monolithic} denote the two references methods introduced earlier. For all three 360$^\circ$ video streaming systems under comparison, we measured the video quality per viewport experienced by a user navigating the 360$^\circ$ content, as the luminance PSNR of the MSE of the pixels displayed in that viewport.
\begin{figure}[htb]
\centering
\vspace{-0.2cm}
\includegraphics[width=0.9\columnwidth]{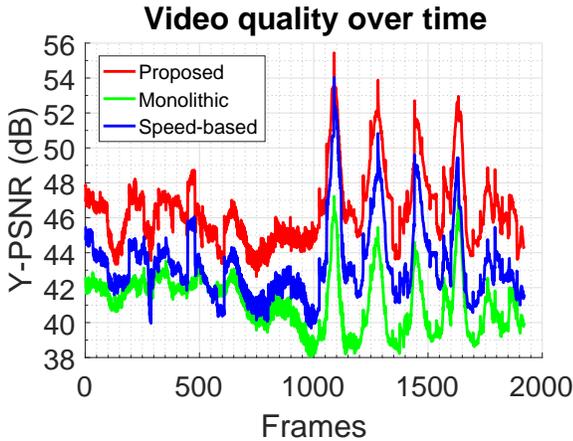}
\vspace{-0.1cm}
\caption{360$^\circ$ viewport video quality: Roller Coaster.}
\label{QoE}
\vspace{-0.3cm}
\end{figure}
Figure~\ref{QoE} shows the viewport PSNR over time for the three competing systems in the case of the Roller Coaster video. Here, we encoded the monolithic 360$^\circ$ video ({\em Monolithic}) with a fixed QP value of 36 and recorded the resulting data rate per GOP to use it as the corresponding rate constraint in \eqref{eqn:Optimization} for our optimization ({\em Proposed}) and similarly for the other reference method ({\em Speed-based}). We can see from Figure~\ref{QoE} that all three systems exhibit the same temporal pattern in viewport PSNR variations, as the dynamic 360$^\circ$ content evolves, with our framework outperforming the two reference methods consistently and considerably. We also observed that {\em Speed-based} offers an improved performance over {\em Monolithic}, when viewport prediction succeeds. Though there are minor variations for some frames, we observed that on average {\em Proposed} provides a 5 dB gain over {\em Monolithic} and a 3 dB gain over {\em Speed-based}.

\begin{figure}[htb]
\centering
\vspace{-0.2cm}
\includegraphics[width=0.9\columnwidth]{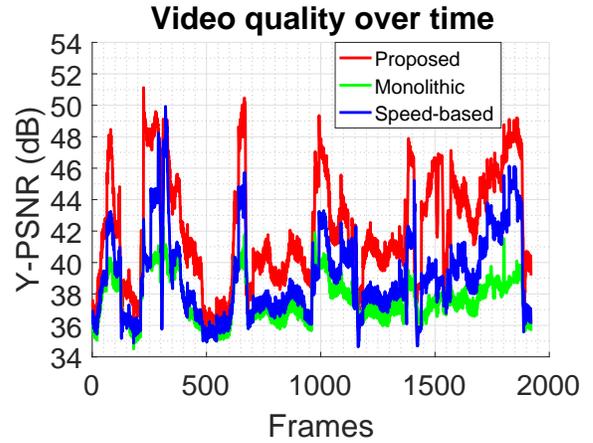}
\vspace{-0.1cm}
\caption{360$^\circ$ viewport video quality: Wingsuit.}
\vspace{-0.2cm}
\label{QoE2}
\end{figure}

We observe a different viewport PSNR pattern for the Wingsuit video, as seen from Figure~\ref{QoE2}. As noted in Figure~\ref{weight2} earlier, the navigation likelihoods over the applied tiling are more uniform in this case, which means the user viewport varies more over time. This negatively impacts the performance of {\em Speed-based}, which now appears closer to that of {\em Monolithic}. Still, due to its rate-distortion foundation {\em Proposed} outperforms again these two reference methods, enabling an average gain of 3 dB and 4 dB relative to {\em Speed-based} and {\em Monolithic}, respectively.

\begin{figure}[htb]
\centering
\vspace{-0.2cm}
\includegraphics[width=0.9\columnwidth]{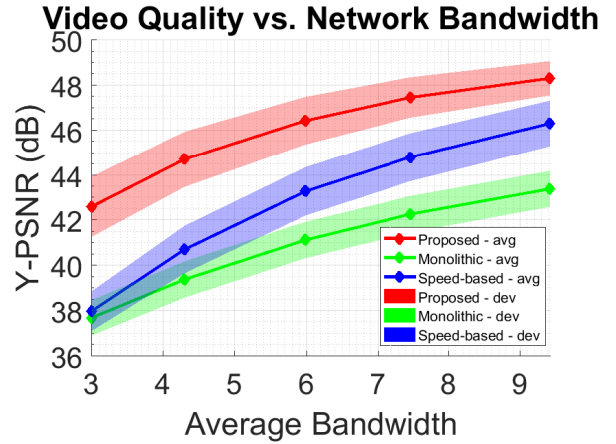}
\vspace{-0.1cm}
\caption{Average 360$^\circ$ viewport video quality: Roller Coaster.}
\vspace{-0.2cm}
\label{QoE_avg}
\end{figure}

Next, we examine the average (over time) viewport 360$^\circ$ video quality (Y-PSNR) delivered by the three competing systems, as the available network bandwidth $C$ is varied. Figure~\ref{QoE_avg} show these results in the case of Roller Coaster, together with the corresponding video quality standard deviation exhibited by each system. We can see that again {\em Proposed} outperforms {\em Speed-based} and {\em Monolithic}, with a consistent gain of up to 4-5 dB, across the entire range of values examined for $C$. As expected, {\em Proposed} and {\em Speed-based} exhibit a somewhat higher viewport video quality standard deviation relative to {\em Monolithic}, since the latter encodes all tiles with a uniform QP value (thus video quality). On the other hand, the reconstruction error can vary more spatially across pixels in viewports delivered by {\em Proposed} and {\em Speed-based}, due to the applied tiling, especially as the number of tiles that comprise a viewport increases.

\begin{figure}[htb]
\centering
\vspace{-0.2cm}
\includegraphics[width=0.9\columnwidth]{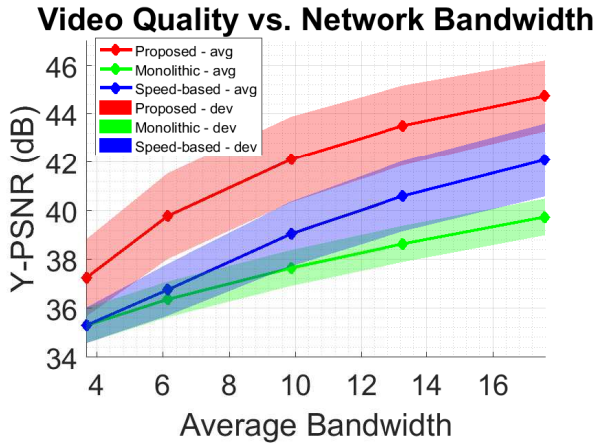}
\caption{Average 360$^\circ$ viewport video quality: Wingsuit.}
\vspace{-0.2cm}
\label{QoE_avg2}
\end{figure}

This phenomenon is emphasized even more in the case of Wingsuit, as indicated by the corresponding results in Figure~\ref{QoE_avg2}, since the navigation likelihoods for tiles closer to the south pole are higher for this 360$^\circ$ video content (see Figure~\ref{weight2}), which in turn causes viewports to more often comprise a higher number of tiles.

We measured in our experiments that our framework leads to 42\% rate savings relative to the conventional approach {\em Monolithic}, which is very encouraging.

%
%CDF of the PSNR values of Roller Coaster video are compared in figure \ref{QoE_cdf}. In the proposed approach, CDF has more gradual increase since PSNR values has a wider deviation than the reference case.
%
%\begin{figure}[htb]
%\centering
%\includegraphics[width=0.9\columnwidth]{Figures/cdf.eps}
%\caption{CDF of PSNR for all frames.}
%\label{QoE_cdf}
%\end{figure}

\section{Conclusion}
\label{Conclusion}
We have formulated a framework for viewport-driven rate optimized $360^\circ$ video streaming that integrates the user view navigation pattern and the spatiotemporal rate-distortion characteristics of the $360^\circ$ video content to maximize the delivered user quality of experience for the given network/system resources. It comprises a methodology for constructing dynamic heat maps that capture the user likelihood of navigating different spatial segments of a $360^\circ$ video over time, analysis and characterization of its spatiotemporal rate-distortion characteristics that leverages preprocessed spatial tilling of the $360^\circ$ view sphere, and optimization problem formulation that characterizes the delivered user quality of experience given the user navigation patterns, $360^\circ$ video encoding decisions, and the available system/network resources. Our experimental results demonstrate the advantages of our framework over the conventional approach of streaming a monolithic uniformly-encoded $360^\circ$ video and a state-of-the-art reference method, enabling considerable video quality of gains of 4 - 5 dB in the case of two popular 4K $360^\circ$ videos.

There are multiple directions of future work that we consider. In the present framework, we used a given tiling of the $360^\circ$ view panorama. How the end-to-end performance efficiency varies with the employed $360^\circ$ tiling is one question we plan to answer. Similarly, will variable-size $360^\circ$ tiling provide additional gains, and at what cost, is another question we will aim to investigate. Finally, we plan to explore adaptive scalable $360^\circ$ tiling representations that will account for client and network heterogeneity intrinsically. Leveraging them towards the design of effective $360^\circ$ network multicast techniques is another study we plan to carry out in this context.

\addtolength{\textheight}{-5cm}   % This command serves to balance the column lengths

%%%%%%%%%%%%%%%%%%%%%%%%%%%%%%%%%%%%%%%%%%%%%%%%%%%%%%%%%%%%%%%%%%%%%%%%%%%%%%%%

%%%%%%%%%%%%%%%%%%%%%%%%%%%%%%%%%%%%%%%%%%%%%%%%%%%%%%%%%%%%%%%%%%%%%%%%%%%%%%%%

%%%%%%%%%%%%%%%%%%%%%%%%%%%%%%%%%%%%%%%%%%%%%%%%%%%%%%%%%%%%%%%%%%%%%%%%%%%%%%%%

\bibliographystyle{IEEEtran}
\bibliography{bibfile}

\begin{thebibliography}{10}
\providecommand{\url}[1]{#1}
\csname url@samestyle\endcsname
\providecommand{\newblock}{\relax}
\providecommand{\bibinfo}[2]{#2}
\providecommand{\BIBentrySTDinterwordspacing}{\spaceskip=0pt\relax}
\providecommand{\BIBentryALTinterwordstretchfactor}{4}
\providecommand{\BIBentryALTinterwordspacing}{\spaceskip=\fontdimen2\font plus
\BIBentryALTinterwordstretchfactor\fontdimen3\font minus
  \fontdimen4\font\relax}
\providecommand{\BIBforeignlanguage}[2]{{%
\expandafter\ifx\csname l@#1\endcsname\relax
\typeout{** WARNING: IEEEtran.bst: No hyphenation pattern has been}%
\typeout{** loaded for the language `#1'. Using the pattern for}%
\typeout{** the default language instead.}%
\else
\language=\csname l@#1\endcsname
\fi
#2}}
\providecommand{\BIBdecl}{\relax}
\BIBdecl

{\footnotesize

\bibitem{digi2017}
\BIBentryALTinterwordspacing
T.~Merel. (2017, Jan.) After a mixed year, mobile {AR} to drive \$108 billion
  {VR/AR} market by 2021. [Online]. Available: \url{http://goo.gl/Lxf4Sy}
\BIBentrySTDinterwordspacing

\bibitem{ApostolopoulosCCKTW:12}
J.~G. Apostolopoulos, P.~A. Chou, B.~Culbertson, T.~Kalker, M.~D. Trott, and
  S.~Wee, ``The road to immersive communication,'' \emph{Proceedings of the
  IEEE}, vol. 100, no.~4, pp. 974--990, Apr. 2012.

\bibitem{Corbillon2017}
X.~Corbillon, G.~Simon, A.~Devlic, and J.~Chakareski, ``Viewport-adaptive
  navigable 360-degree video delivery,'' in \emph{Proc. IEEE Int'l
  Conf. Communications (ICC)}, Paris, France, May 2017.

\bibitem{Yu2015}
M.~Yu, H.~Lakshman, and B.~Girod, ``A framework to evaluate omnidirectional
  video coding schemes,'' in \emph{Proc. IEEE International Symposium on Mixed
  and Augmented Reality}, Sep. 2015, pp. 31--36.

\bibitem{Facebook360}
\BIBentryALTinterwordspacing
``Facebook 360: A stunning and captivating way to share immersive stories,
  places and experiences.''
  \url{http://facebook360.fb.com}
\BIBentrySTDinterwordspacing

\bibitem{YouTube360}
\BIBentryALTinterwordspacing
``{YouTube: 360$^\circ$ Videos}.''
  \url{https://www.youtube.com/}
\BIBentrySTDinterwordspacing

\bibitem{Forbes_VR_2016}
B.~Begole, ``Why the {Internet} pipes will burst when virtual reality takes
  off,'' Forbes Magazine, Feb. 2016.

\bibitem{EdwardKnightlyKeynoteINFOCOM2017}
E.~Knightly, ``Scaling {Wi-Fi} for next generation transformative
  applications,'' Keynote Presentation, IEEE INFOCOM, Atlanta, GA, May
  2017.

\bibitem{MossM:11}
J.~D. Moss and E.~R. Muth, ``Characteristics of headmounted displays and their
  effects on simulator sickness,'' \emph{Human Factors: The Journal of the
  Human Factors and Ergonomics Society}, Jun.
  2011.

\bibitem{Afzal2017}
S.~Afzal, J.~Chen, and K.~K. Ramakrishnan, ``Characterization of 360-degree
  videos,'' in \emph{Proc. ACM SIGCOMM Workshop Virtual Reality and Augmented
  Reality Network}, Aug. 2017.

\bibitem{SullivanOHW:12}
G.~J. Sullivan, J.-R. Ohm, W.-J. Han, and T.~Wiegand, ``Overview of the high
  efficiency video coding {(HEVC)} standard,'' \emph{IEEE Trans. Circuits and
  Systems for Video Technology}, Dec. 2012.

\bibitem{Ozcinar2017}
C.~Ozcinar, A.~De~Abreu, and A.~Smolic, ``Viewport-aware adaptive 360$^\circ$
  video streaming using tiles for virtual reality,'' in \emph{Proc. IEEE
  Int'l Conf. Image Processing}, Beijing, China, Sep. 2017.

\bibitem{Petrangeli2017}
S.~Petrangeli, V.~Swaminathan, M.~Hosseini, and F.~De~Turck, ``Improving
  virtual reality streaming using {HTTP/2},'' in \emph{Proc. ACM Multimedia Systems
  Conference}, Jun. 2017, pp. 225--228.

\bibitem{Qian2016}
F.~Qian, L.~Ji, B.~Han, and V.~Gopalakrishnan, ``Optimizing 360 video delivery
  over cellular networks,'' in \emph{Proc. ACM Workshop All Things Cellular:
  Operations, Applications and Challenges}, Oct. 2016.

\bibitem{Hosseini2016}
M.~Hosseini and V.~Swaminathan, ``Adaptive 360 {VR} video streaming: Divide and
  conquer,'' in \emph{Proc. IEEE Int'l Symp. Multimedia}, Dec. 2016.

\bibitem{Graf2017}
M.~Graf, C.~Timmerer, and C.~Mueller, ``Towards bandwidth efficient adaptive
  streaming of omnidirectional video over {HTTP}: Design, implementation, and
  evaluation,'' in \emph{Proc. ACM MM Syst. Conf}, Jun. 2017.

\bibitem{Yahia2017}
M.~Ben~Yahia, Y.~Le~Louedec, L.~Nuyami, and G.~Simon, ``When {HTTP/2} rescues
  {DASH}: Video frame multiplexing,'' in \emph{Proc. IEEE INFOCOM Workshop
  Communication and Networking Techniques for Contemporary Video}, Atlanta, GA,
  USA, May 2017.

\bibitem{Nasrabadi2017}
A.~T. Nasrabadi, A.~Mahzari, J.~D. Beshay, and R.~Prakash, ``Adaptive
  360-degree video streaming using scalable video coding,'' in
  \emph{Proc. ACM Multimedia Conference}, 2017, pp. 1689--1697.

\bibitem{He2018}
G.~He, J.~Hu, H.~Jiang, and Y.~Li, ``Scalable video coding based on user's view
  for real-time virtual reality applications,'' \emph{IEEE Communications
  Letters}, vol.~22, no.~1, pp. 25--28, Jan 2018.

\bibitem{Coaster}
\BIBentryALTinterwordspacing
``Mega coaster: Get ready for the drop (360 video).'' [Online]. Available:
  \url{https://youtu.be/-xNN-bJQ4vI}
\BIBentrySTDinterwordspacing

\bibitem{Dubai}
\BIBentryALTinterwordspacing
``Wingsuit 360 degree video over {Dubai}.'' [Online]. Available:
  \url{https://youtu.be/AX4hWfyHr5g}
\BIBentrySTDinterwordspacing

\bibitem{BoydV:03}
S.~Boyd and L.~Vandenberghe, \emph{Convex Optimization}.\hskip 1em plus 0.5em
  minus 0.4em\relax Cambridge University Press, 2004.

\bibitem{OpenTrack}
\BIBentryALTinterwordspacing
``Opentrack: Head tracking software.'' [Online]. Available:
  \url{https://github.com/opentrack/opentrack}
\BIBentrySTDinterwordspacing

}

\end{thebibliography}

\end{document}